\documentclass{article}
\usepackage{srcltx}
\usepackage{amsmath}
\usepackage{amssymb}
\usepackage{amsfonts}
\usepackage{latexsym}
\usepackage{textcomp}
\usepackage{multirow}
\usepackage{booktabs}
\usepackage{url}
\usepackage{array}
\usepackage{marvosym}
\usepackage{graphics}
\usepackage{graphicx}
\usepackage[caption=false]{subfig}
\usepackage{adjustbox}
\usepackage{epsfig}
\usepackage[square,numbers]{natbib}

\usepackage{lineno}

\title{Some stylized facts of the Bitcoin market}

\author{Aurelio F. Bariviera \\ 
{\tiny aurelio.fernandez@urv.cat} \and
Mar\'ia Jos\'e Basgall \\
{\tiny mjbasgall@lidi.info.unlp.edu.ar}
 \and Waldo Hasperu\'e \\
{\tiny whasperue@lidi.info.unlp.edu.ar}
\and Marcelo Naiouf \\
{\tiny mnaiouf@lidi.info.unlp.edu.ar}
}

\begin{document}

\maketitle

\begin{abstract}
In recent years a new type of tradable assets appeared, generically known as cryptocurrencies. Among them, the most widespread is Bitcoin. Given its novelty, this paper investigates some statistical properties of the Bitcoin market. This study compares Bitcoin and standard currencies dynamics and focuses on the analysis of returns at different time scales. We test the presence of long memory in return time series from 2011 to 2017, using transaction data from one Bitcoin platform. We compute the Hurst exponent by means of the Detrended Fluctuation Analysis method, using a sliding window in order to measure long range dependence. We detect that Hurst exponents changes significantly during the first years of existence of Bitcoin, tending to stabilize in recent times. Additionally, multiscale analysis shows a similar behavior of the Hurst exponent, implying a self-similar process. 
 \end{abstract}

\section{Introduction \label{sec:intro}}

According to the traditional definition, a currency has three main properties: (i) it serves as a medium of exchange, (ii) it is used as a unit of account and (iii) it allows to store value. Along economic history, monies were related to political power. In the beginning, coins were minted in precious metals. Therefore, the value of a coin was intrinsically determined by the value of the metal itself. Later, money was printed in paper bank notes, but its value was linked somewhat to a quantity in gold, guarded in the vault of a central bank. Nation states have been using their political power to regulate the use of currencies and impose one currency (usually the one issued by the same nation state) as legal tender for obligations within their territory. In the twentieth century, a major change took place: abandoning gold standard. The detachment of the currencies (specially the US dollar) from the gold standard meant a recognition that the value of a currency (specially in a world of fractional banking) was not related to its content or representation in gold, but to a broader concept as the confidence in the economy in which such currency is based. In this moment, the value of a currency reflects the best judgment about the monetary policy and the ``health'' of its economy.
 
\begin{figure}
\center \includegraphics[scale=.6]{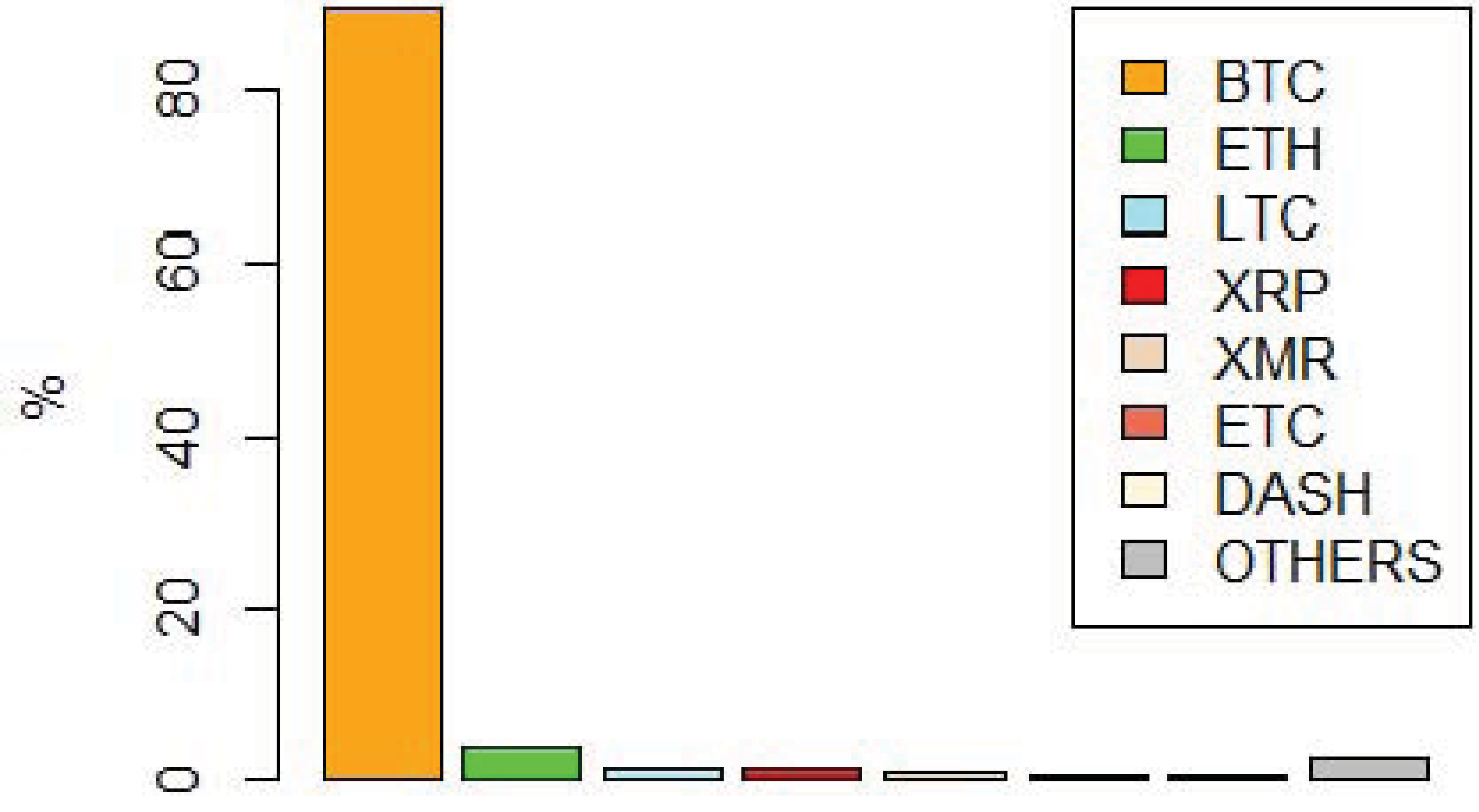}
\caption{Cryptocurrencies. Share of market capitalization of each currency. Own elaboration based on data from \cite{coinmarketcap}}
\label{cryptocapitalzation}
\end{figure}

In recent years, a new type of currencies, a synthetic one, emerged. We name this new type as ``synthetic'' because it is not the decision of a nation state, nor represents any underlying asset or tangible wealth source.  It appears as a new tradable asset resulting from a private agreement and facilitated by the anonymity of internet. Among this synthetic currencies, Bitcoin (BTC) emerges as the most important one, with a market capitalization of 15 billions, as of December 2016. There are other cryptocurrencies, based on blockchain technology, such as Litecoin (LTC), Ethereum (ETH), Ripple (XRP). The website \texttt{https://coinmarketcap.com/currencies/} counts up to 641 of such monies. However, as we can observe in Figure \ref{cryptocapitalzation}, Bitcoin represents 89\% of the capitalization of the market of all cryptocurrencies. One open question today is if Bitcoin is in fact a, or may be considered as a, currency. Until now, we cannot observe that Bitcoin fulfills the main properties of a standard currency. It is barely accepted as a medium of exchange (e.g. to buy some products online), it is not used as unit of account (there are no financial statements valued in Bitcoins), and we can hardly believe that, given the great swings in price, anyone can consider Bitcoin as a suitable option to store value. Given these characteristics, Bitcoin could fit as an ideal asset for speculative purposes. There is no underlying asset to relate its value to and there is an open platform to operate round the clock. 

The aim of this paper is to study some statistical characteristics of Bitcoin \textit{et al.}{vis-\`a-vis} some major currencies, during the period 2011-2017. We will focus our attention on the evolution of the long memory of the time series.  This article contributes to the literature  in three important aspects. First, we expand the empirical studies by analyzing the long memory of a new asset. Second, we compare the behavior of Bitcoin with some major currencies. Third, we highlight the evolution in the underlying dynamics of this new  market. The rest of the paper is organized as follows: Section \ref{sec:literature} describes the recent emerging literature on Bitcoin, Section \ref{sec:methodology} describes the methodology used in the paper, Section \ref{sec:data_results} presents the data and results of our empirical analysis and, finally Section \ref{sec:conclusions} draws the main conclusions.

\section{Brief literature review\label{sec:literature}}
\subsection{Bitcoin}
Speculation has a long history and it seems inherent to capitalism. One common feature of speculative assets in history has been the difficulty in valuation. Tuplipmania, the South Sea bubble, and more others, reflect on one side human greedy behavior, and on the other side, the difficulty to set an objective value to an asset. All speculative behaviors were reflected in a super-exponential growth of the time series \cite{FeigenbaumFreund1996}.

Cryptocurrencies can be seen as the libertarian response to central bank failure to manage financial crises, as the one occurred in 2008. Also cryptocurrencies can bypass national restrictions to international transfers, probably at a cheaper cost.
Bitcoin was created by a person or group of persons under the pseudonym Satoshi Nakamoto. The description of Bitcoin Core, i.e. the open source client of the Bitcoin cryptocurrency, is described in \cite{Nakamoto}

The discussion of Bitcoin has several perspectives. The computer science perspective deals with the strengths and weaknesses of blockchain technology. In fact, according to \cite{BoEngland}, the introduction of a ``distributed ledger'' is the key innovation. Traditional means of payments (e.g. a credit card), rely on a central clearing house that validate operations, acting as ``middleman'' between buyer and seller. On contrary, the payment validation system of Bitcoin is decentralized. There is a growing army of miners, who put their computer power at disposal of the network, validating transactions by gathering together blocks, adding them to the ledger and forming a 'block chain'.  This work is remunerated by giving the miners Bitcoins, what makes (until now) the validating costs cheaper than in a centralized system. The validation is made by solving some kind of algorithm. With the time solving the algorithm becomes harder, since the whole ledger must be validated. Consequently it takes more time to solve it. Contrary to traditional currencies, the total number of Bitcoins to be issued is beforehand fixed: 21 million. In fact, the issuance rate of Bitcoins is expected to diminish over time. According to \cite{VCdenmark}, validating the public ledger was initially rewarded with 50 Bitcoins, but the protocol forsee halving this quantity every four years. At the current pace, the maximum number of Bitcoins will be reached in 2140. Taking into account the decentralized character, Bitcoin transactions seem secure. All transactions are recorded in several computer servers around the world. In order to commit fraud, a person should change and validate (simultaneously) several ledgers, which is almost impossible. Additional, ledgers are public, with encrypted identities of parties, making transactions ``pseudonymous, not anonymous'' \cite{Congressional}. 

The legal perspective of Bitcoin is fuzzy. Bitcoin is not issued, nor endorsed by a nation state. It is not an illegal substance. As such, its transaction is not regulated. 

The economic perspective is still under study. The use of Bitcoin in daily life is marginal. At the time of writing this paper, there were only 8367 retailers worldwide who accepted Bitcoins as a means of payment, mostly concentrated in North America, western Europe, and some major cities in South America and South East Asia \cite{coinmap}. There is not too much information regarding Bitcoin exchanges. This gray situation raises some concerns about a possible Ponzi scheme. There are no savings accounts in Bitcoins and consequently no interest rates. All these elements together contribute to its difficulty to assess a fair value. Cheung \textit{et al.} \cite{CheungRoca2015} detect several price bubbles over the period 2010-2014. Three of them lasted from 66 to 106 days to burst. Ciaian and coworkers \cite{Ciaian2016} find no macro-financial indications driving Bitcoin price, and they do not discard that investor speculation affects significantly the price evolution.

\subsection{The Efficient Market Hypothesis}

As recalled in the previous section, the nature of the Bitcoin is not yet clear. In particular, given the nonexistence of saving accounts in Bitcoin, and consequently the absense of a Bitcoin interest rate, precludes the idea of studying the price behavior in relation with cash flows generated by Bitcoins.
As a consequence, we aim to analize the underlying dynamics of the price signal, using the Efficient Market Hypothesis as a theoretical framework.
The Efficient Market Hypothesis (EMH) is the cornerstone of financial economics. One of the seminal works on the stochastic dynamics of speculative prices is due to Bachelier \cite{Bachel}, who in his doctoral thesis developed the first mathematical model concerning the behavior of stock prices. The systematic study of informational efficiency begun in the 1960s, when financial economics was born as a new area within economics. The classical definition due to Eugene Fama \cite{Fama76} says that a market is informationally efficient if it ``fully reflect all available information''. Therefore, the key element in assessing efficiency is to determine the appropriate set of information that impels prices. Following \cite{Fama70}, informational efficiency can be divided into three categories: (i) weak efficiency, if prices reflect the information contained in the past series of prices, (ii) semi-strong efficiency, if prices reflect all public information and (iii) strong efficiency, if prices reflect all public and private information. As a corollary of the EMH, one cannot accept the presence of long memory in financial time series, since its existence would allow a riskless profitable trading strategy. If markets are informationally efficient, arbitrage prevent the possibility of such strategies. 

An important part of the literature focused its attention on studying the long-range dependence. If we consider the financial market as a dynamical structure, short term memory can exist (to some extent) without contradicting the EMH. In fact, the presence of some mispriced assets is the necessary stimulus for individuals to trade and reached an (almost) arbitrage free situation. However, the presence of long range memory is at odds with the EMH, because it would allow an stable trading rule to beat the market. 

Given the novelty of Bitcoin, this is one of the first papers (probably with the single exception of \cite{Kristoufek201627}) to study the Hurst exponent of this market. Previous works on long range dependence focused their attention in stocks, bonds or commodities markets.
In particular, \cite{GreeneFielitz77} and \cite{Mills93} use the Hurst exponent to detect the presence of long memory in the US and the UK stock markets, respectively. In \cite{FF88} positive short term autocorrelation and negative long term autocorrelation is found, after examining the returns of a diversified portfolio of the NYSE. This result reinforces the idea of an underlying mean-reverting process. Long memory is also found in the Spanish stock market \cite{BlascoSantamaria96} and the Turkish stock market \cite{Kilic04}. In the same line, Barkoulas \textit{et al.} \cite{BarkoulasBaumTravlos00} finds evidence of long memory in the weekly returns of the Athens Stock Exchange during the period 1981-1990, and suggest that the strength of the memory could be influenced by the market size. Also long memory behavior in the Greek market was found by Panas \cite{Panas01}. Cajueiro and Tabak \cite{CajueiroTabak04} find that developed markets are more informationally efficient than emerging markets and that the level of efficiency is influenced by market size and trading costs. Cajueiro and Tabak \cite{CajueiroTabak05causes} relate long-range dependence with specific financial variables of the firms under examinations. Zunino and coworkers \cite{ZuninoPhyB07} find that the long-range memory in seven Latin-American markets is time varying. In this line, Bariviera \cite{Bariviera2011} finds evidence of a time varying long-range dependence in daily returns of Thai Stock Market during the period 1975-2010 and concludes that it is weakly influenced by the liquidity level and market size. Vodenska \textit{et al.} \cite{Vodenskaetal08} show that volatility clustering in the S \& P 500 index produces memory in returns. \cite{LaSpada08} finds long memory in the sign of transactions but not in the signs of returns. Ureche-Rangau and  de Rorthays,\cite{UrecheRangau09} investigate the presence of long memory in volatility and trading volume of the Chinese stock market. Cajueiro and Tabak \cite{CajueiroTabakUS} present empirical evidence of time-varying long-range dependence for US interest rates. It concludes that long memory has reduced over time. Moreover, Cajueiro and Tabak\cite{CajueiroTabak2010} find that this long-range dependence, is affected by the monetary policy. Similarly, Cajueiro and Tabak \cite{CajueiroTabakTSIR} find long range dependence in Brazilian interest rates and their volatility, providing important implications for monetary studies. Time-varying long range dependence in Libor interest rates is found in \cite{EPJB2015,BarivieraRSTA2015}. The authors conclude that such behavior is consistent with the Libor rate rigging scandal.

Cheung and Lai \cite{CheungLai95} use the fractional differencing test for long memory by \cite{GPH83} and find evidence of long memory in 5 out of the 18 markets under study. Using a different methodology, \cite{BarkoulasBaum96} applies spectral regression to time series of 30 firms, 7 sector indices and 2 broad stock indices at daily and monthly frequency, and finds evidence of long memory only in 5 of the individual firms. Wright \cite{Wright01} compares the memory content of the time series in developed and emerging stock markets, finding that the latter exhibits short term serial correlation in addition to long-range memory.  Henry \cite{Henry02} concludes that there is strong evidence of long-range memory in the Korean market and some weak evidence on the German, Japanese and Taiwanese markets, after analyzing monthly returns of nine stock markets. Also, Tolvi \cite{Tolvi03} uses a sample of 16 stock markets of OECD countries and finds evidence of long memory only in 3 of them and Kasman \textit{et al.} \cite{KasmanTurgutAyhan09} finds that among the four main central European countries (Czech Republic, Hungary, Poland and Slovak Republic), only the last one exhibits long memory. Cheong \cite{Cheong2010} computes the Hurst exponent by means of three heuristic methods and find evidence of long memory in the returns of five Malaysian equity market indices. This study finds that the Asian economic crisis affected the extent of long-range memory of the Malaysian stock market.

With respect to the fixed income market, Carbone \cite{Carbone04} finds local variability of the correlation exponent in the German stock and sovereign bond markets. Bariviera \textit{et al.}.  \cite{BaGuMa12} finds empirical evidence of long memory in corporate and sovereign bond markets and detects that the current financial crisis affects more the informational efficiency of the corporate than sovereign market. Zunino \textit{et al.}. \cite{Zunino2012}, using the complexity-entropy causality plane for a sample of thirty countries, finds that informational efficiency is related to the degree of economic development. Recently, Bariviera \textit{et al.}. \cite{BaGuMa14} finds that the long range memory of corporate bonds at European level are affected unevenly during the financial crisis. In particular, sectors closely related to financial activities were the first to exhibit a reduction in the informational efficiency.

There are some works that find no evidence of long memory in the financial time series. Among others we can cite Lo \cite{Lo91}, in the returns of US stocks, and Grau-Carles \cite{GrauCarles05} in the stock indices of US, UK, Japan and Spain.

As we can appreciate, the empirical studies on sovereign and corporate bond markets and stock markets are abundant. Giving the increasing amounts involved in Bitcoin trading, we believe that this topic deserves a detailed study.

\section{Long range dependence \label{sec:methodology}}
The presence of long range dependence in financial time series generates a vivid debate. Whereas the presence of short term memory can stimulate investors to exploit small extra returns, making them disappear, long range correlations poses a challenge to the established financial model. As recognized by \cite{Ciaian2016}, Bitcoin price is not driven by macro-financial indicators. Consequently a detailed analysis of the underlying dynamics becomes important to understand its emerging behavior.

There are several methods (both parametric and non parametric) to calculate the Hurst exponent. For a survey on the different methods for estimating long range dependences see \cite{Taqqu95} and \cite{Montanari99}. Serinaldi \cite{Serinaldi10} makes a critical review on the different estimation methods of the Hurst exponent, concluding that an inappropriate application of the estimation method could lead to incorrect conclusions about the persistence or anti-persistence of financial series. Although $R/S$ method is probably one of the most extended methods to approximate long run memory in time series, it is not robust to departures from stationarity. Consequently, if the process under scrutiny exhibits short memory, the $R/S$ statistic could indicate erroneously the presence of long memory. In this sense, \cite{Mosaic94} develops the method called Detrended Fluctuation Analysis (DFA) that is more appropriate when dealing with nonstationary data. As recognized by \cite{GrauCarles}, this method avoids spurious detection of long-range dependence due to nonstationary data. Due to this reason we select the DFA method in order to assess the existence of long memory in this paper.

The algorithm, described in detail in \cite{Peng95}, begins by computing the mean of the stochastic time series $y(t)$, for $t=1,\dots, M$. Then, an integrated time series $x(i)$, $i=1,\dots, M$ is obtained by subtracting mean and adding up to the $i-th$ element, $x(i)=\sum_{t=1}^{i}[y(t)-\bar{y}]$. Then $x(i)$ is divided into $M/m$ non overlapping subsamples and a polynomial fit $x_{pol}(i,m)$ is computed in order to determine the local trend of each subsample. Next the fluctuation function 
\begin{equation}
F(m)=\sqrt{\frac{1}{M}\sum_{i=1}^{M}{[x(i)-x_{pol}(i,m)]}^2}
\label{eq:DFA}
\end{equation}
is computed. This procedure is repeated for several values of $m$. The fluctuation function $F(m)$ behaves as a power-law of $m$, $F(m) \propto m^H$, where $H$ is the Hurst exponent. Consequently, the exponent is computed by regressing $\ln(F(m))$ onto $\ln(m)$. According to the literature the maximum block size to use in partitioning the data is $(length(window)/2)$, where \textit{window} is the time series window vector. Consequently, in this paper we use six points to estimate the Hurst exponent. The points for regression estimation are: $m=\{4, 8, 16, 32, 64, 128\}$. 

There are other methodologies to verify the presence of long-range memory. Rosso \textit{et al.} \cite{Rosso07} introduces the complexity-causality plane in order to discriminate between Gaussian from non-Gaussian processes.  Zunino \textit{et al.} \cite{ZuninoCausality10} shows that this innovative approach could be used to rank stock markets according to their stage of development. In Zunino \textit{et al.}. \cite{ZuninoPermutation11}, the application of the complexity-entropy causality plane was extended to the study of the efficiency of commodity prices. This method reveals that it is not only useful to produce a ranking of efficiency of different commodities, but it also allows to identify periods of increasing and decreasing randomness in the price dynamics. Zunino \textit{et al.} \cite{Zunino2012} uses this representation space to establish an efficiency ranking of different markets and distinguish different bond market dynamics and concludes that the classification derived from the complexity-entropy causality plane is consistent with the qualifications assigned to sovereign instruments by major rating companies.

\section{Data and results \label{sec:data_results}}
The period under study goes from 2011 until 2017 for daily data and from 2013 until 2016 for intraday data. We downloaded the daily prices of Bitcoin and exchange rates of Euro and Sterling Pound, in US dollars. These daily data were downloaded from Datastream. Additionally, we downloaded Bitcoin intraday transaction data from Bitcoin charts website \cite{bitcoincharts}. The original dataset comprises a total of 9540332 transactions. Given that transactions take place irregularly in time, we sampled data each \{5, 6,\dots, 12\} hours. The minimum sample space corresponds to the maximum time without transactions in our dataset. 

We compute the instantaneous return, measured a $r_t=\log(P_t)-\log(P_{t-1})$. With this values we calculate the Hurst exponent using DFA method. In order to assess the change in time in long range memory, following \cite{CajuTabEM,CajuTab}, we use sliding windows. We estimate the Hurst exponent using two year sliding windows (500 datapoints). In particular, we use overlapping windows, moving forward by 1 datapoint, in order to allow for smooth transitions. 

\subsection{Daily returns}
Our first analysis focuses on the descriptive statistics of daily returns of Bitcoin (BTC) \textit{vis-\`a-vis} two major currencies such as Euro (EUR) and the British Pound (GBP). Results are presented in Table \ref{tab:main-currencies-statistics}. Whereas EUR and GBP exhibit similar mean, median and standard deviation values, BTC presents a significant positive mean and median. Moreover, BTC standard deviation  is 10 times greater than of the other currencies. All three currencies are clearly non-normal according to the Jarque-Bera test \cite{JarqueBera1987}.

\begin{table}[htbp]
  \centering
  \caption{Descriptive statistics of daily returns of BTC, EUR and GBP, from 2011 until 2017}
    \begin{tabular}{rrrr}
    \toprule
          & \multicolumn{1}{c}{GBP} & \multicolumn{1}{c}{EUR} & \multicolumn{1}{c}{BTC} \\
    \midrule
    Observations & 1404  & 1404  & 1404 \\
    Mean  & 0.0205 & 0.0219 & 0.3172 \\
    Median & 0.0000 & 0.0033 & 0.2151 \\
    Std Deviation & 0.5701 & 0.5731 & 6.2416 \\
    Skewness & 2.2166 & -0.0418 & -1.1775 \\
    Kurtosis & 36.1865 & 4.8014 & 25.5677 \\
    Jarque Bera & 65578.4593 & 190.2491 & 30118.6642 \\
    \bottomrule
    \end{tabular}%
  \label{tab:main-currencies-statistics}%
\end{table}%

\begin{figure}
\center \includegraphics[scale=.6]{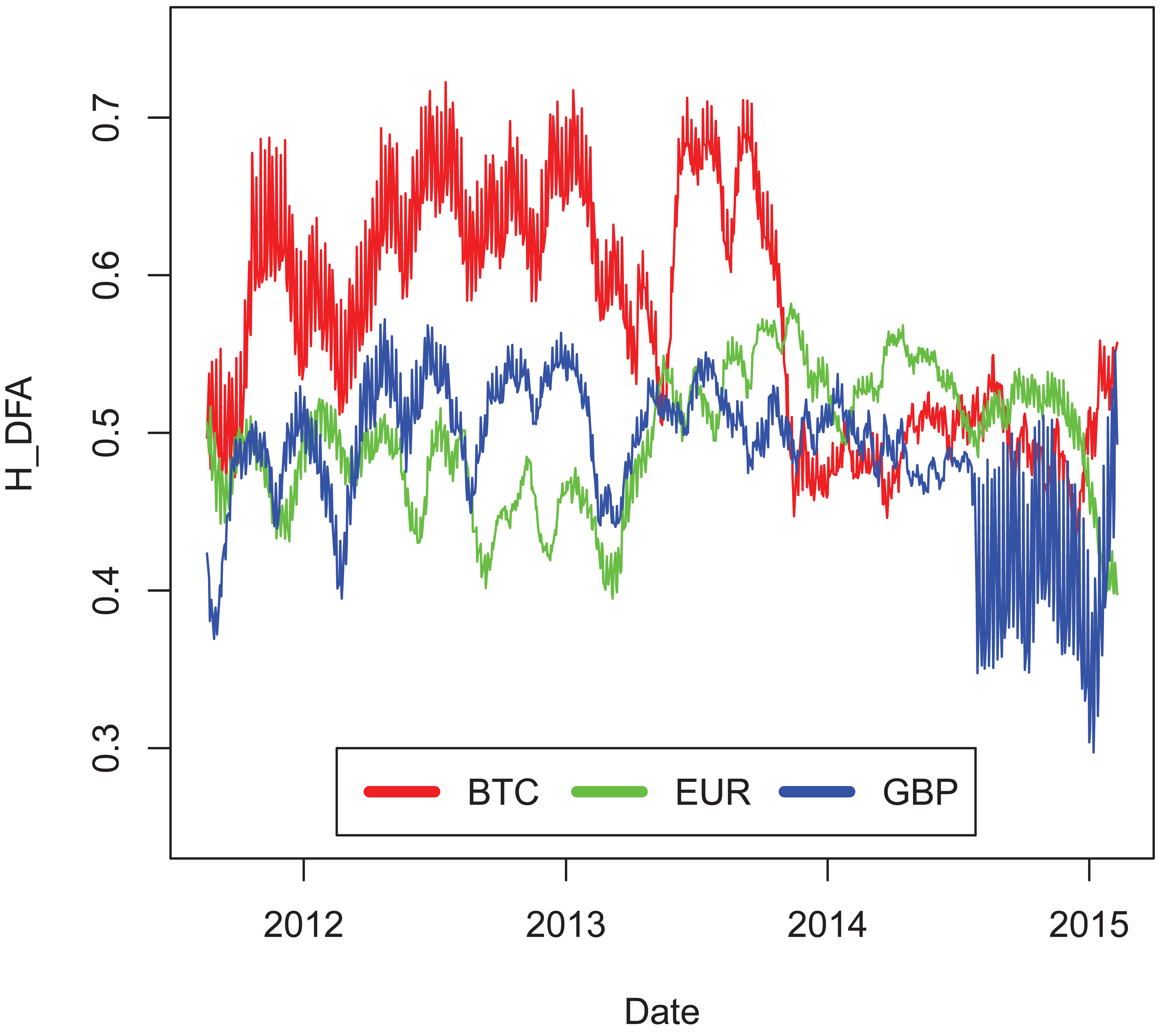}
\caption{Hurst exponent of BTC, EUR and GBP daily values, using a sliding window of 500 datapoints and stepping forward 1 datapoint.}
\label{dfa-main-currencies}
\end{figure}

We continue our analysis computing the long-range memory of all three assets using the DFA method. Figure \ref{dfa-main-currencies}, shows important difference with respect to the stochastic behavior of all three assess. On the one hand, EUR and GBP wanders roughly within the interval $H=(0.45,0.55)$, which reflects an approximate random walk behavior. Except for the last period in GBP, we can say that both currencies behaves accordingly the Efficient Market Hypothesis. Taking into account that both are very liquid markets, we can expect such behavior. On the other hand, BTC returns exhibits long range correlations for most of the period under study. The convergence in memory behavior begins in 2014, where all three currencies meets around $H=0.5$.

\begin{figure}
\center \includegraphics[scale=.4]{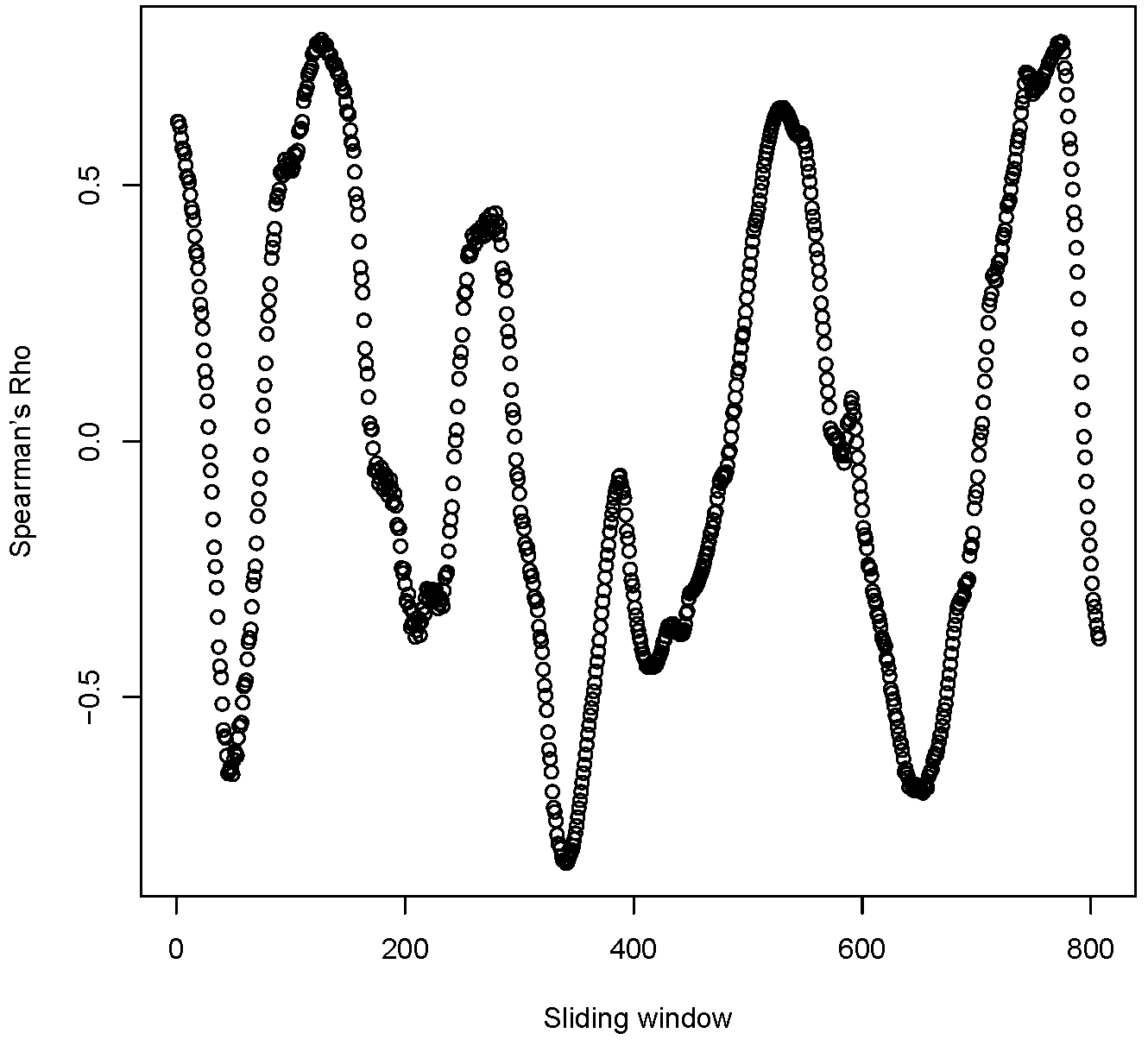}
\caption{Spearman's Rho between Hurst exponent and turnover by volume of BTC.}
\label{spearmanrho}
\end{figure}

We test if the Hurst exponent is, specially in recent times, related to the liquidity level of the market. In order to do so, we run the Spearman's non parametric test, to assess the association between the Hurst exponent and BTC turnover by volume. If we consider the whole period, there is no significant association between both variables. However, if we study this association over time, we observe a time-varying relationship. This situation (see Figure \ref{spearmanrho}) could reflect a detachment of the underlying dynamics from one important market liquidity indicator.
 
\subsection{Intraday returns}
Taking into account that one of the advantages of Bitcoin is its open source philosophy, there is much available data, in order to analyze. Consequently we obtained transaction data from the 31th March 2013 to 2nd August 2016, and we sampled it in order to generate returns by hours, with the aim of disecting the behavior at different time scales.

\begin{figure}
\center \includegraphics[scale=.4]{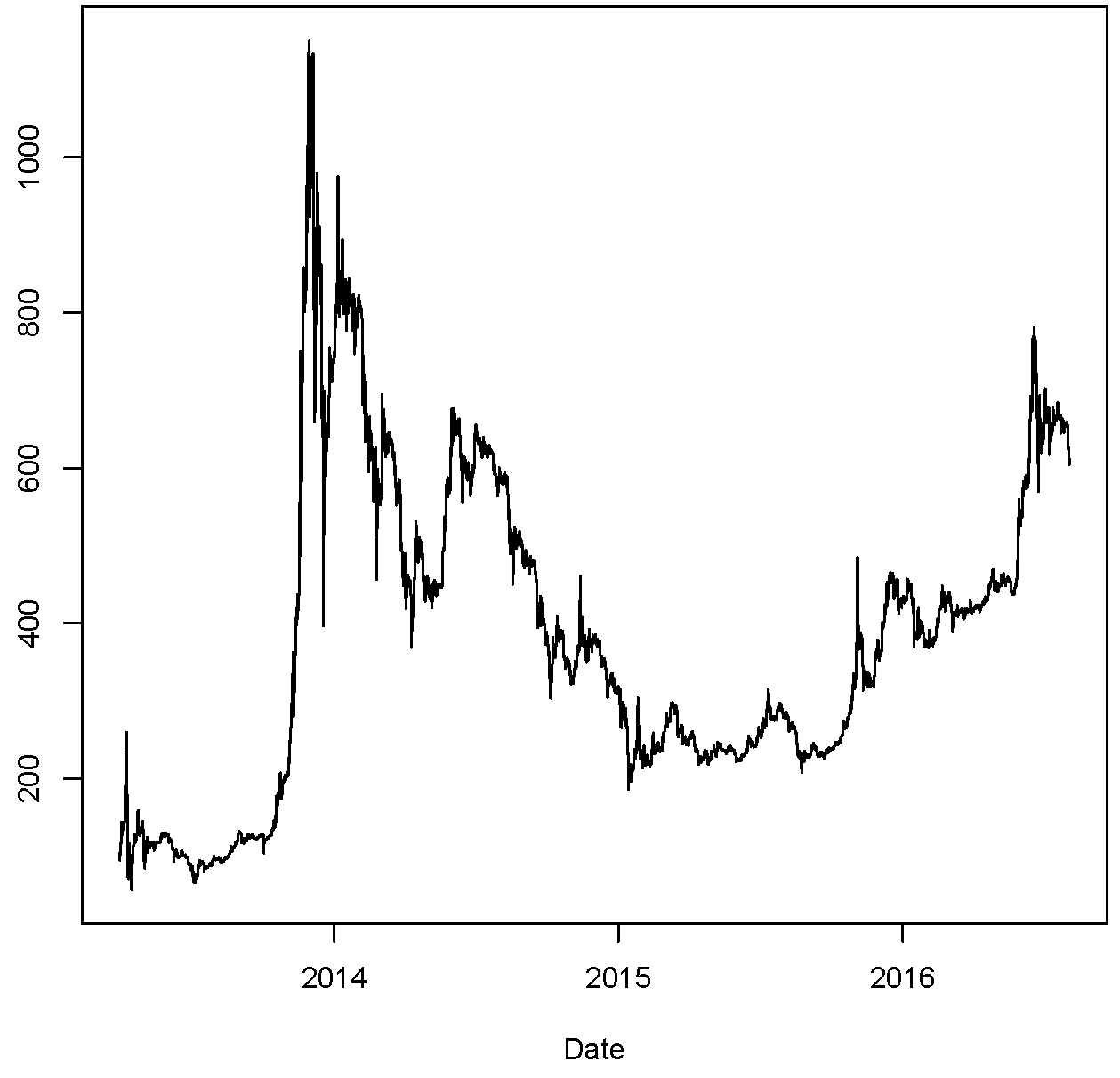}
\caption{Bitcoin price in USD, sampled every 5 hours.}
\label{bitprice}
\end{figure}

In Figure \ref{bitprice}, we appreciate the sometimes meteoric runs-up and down of price. In less than a year, between 2013 and 2014, the price rocketed from less than 100 USD to more than 1000 USD, followed by a several falls and rebounds, without reaching an stability zone.  

\begin{figure}
\center \includegraphics[scale=.4]{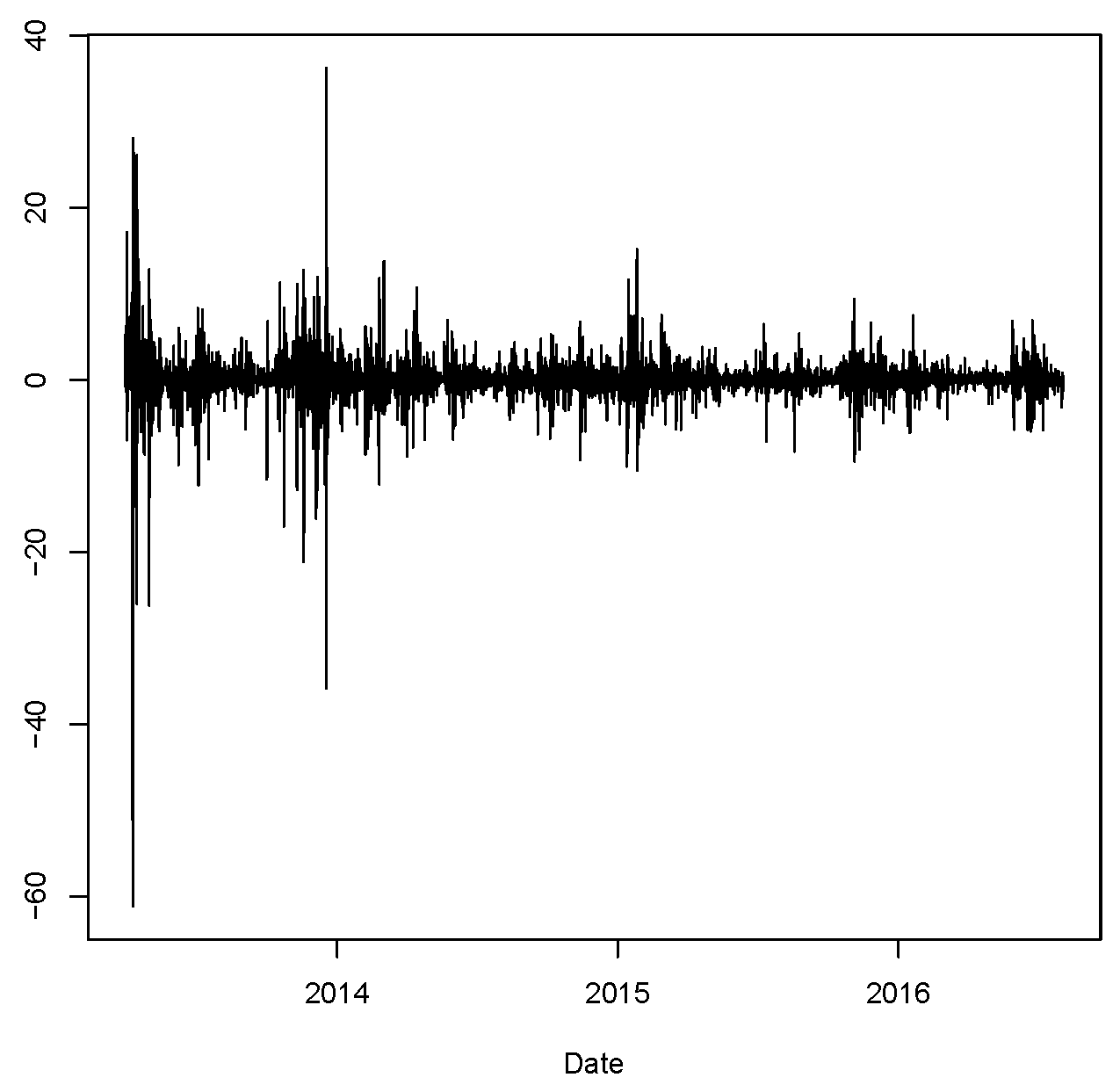}
\caption{Bitcoin returns, sampled every 5 hours.}
\label{bitreturn5h}
\end{figure}

Another aspect we detect is that price volatility (sample variance) shows a diminishing trend. This situation is reflected in Figure \ref{bitreturn5h}.

Table \ref{tab:returnstat} shows the descriptive statistics of Bitcoin returns, for each of the sampling intervals. We observe that, whereas the mean return increases \textit{pari passu} with the interval length, the median return remains around 0.03. Another feature about returns is that they exhibit huge volatility, either measured by the standard deviation or the return range (max-min). In particular, large range values are reflected in the presence of great swings in returns, which can be observed in Figure \ref{bitreturn5h}. Finally, we detect that data is negatively skewed and present an acute excess of kurtosis, which lead to a rejection of the null hypothesis of normality according to the Jarque-Bera statistic. Skewness and kurtosis seem to reduce with greater time spans, which could reflect a slow trend toward a more Gaussian behavior. 
  
\begin{table}[htbp]
  \centering
  \caption{Descriptive statistics of returns, sampled at different time spans}
  \resizebox{\columnwidth}{!}{%
    \begin{tabular}{rrrrrrrrr}
    \toprule
          & 5h    & 6h    & 7h    & 8h    & 9h    & 10h   & 11h   & 12h \\
    \midrule
    Length & 5746  & 4879  & 4182  & 3659  & 3252  & 2927  & 2661  & 2439 \\
    Mean  & 0.0325 & 0.0382 & 0.0445 & 0.0508 & 0.0572 & 0.0632 & 0.0695 & 0.0751 \\
    Median & 0.0359 & 0.0252 & 0.0323 & 0.0246 & 0.0395 & 0.0302 & 0.0235 & 0.0630 \\
    Min   & -61.1397 & -46.4425 & -61.1258 & -40.1405 & -50.4934 & -63.3724 & -40.5581 & -53.6354 \\
    Max   & 36.2219 & 40.3414 & 46.7465 & 48.5574 & 47.7417 & 47.5930 & 29.8259 & 51.3806 \\
    Std. Dev. & 2.5994 & 2.6907 & 3.0265 & 3.2340 & 3.1859 & 3.6885 & 3.4752 & 3.9545 \\
    Skewness & -3.6037 & -2.0001 & -2.9456 & -1.1589 & -1.3924 & -1.8665 & -1.2430 & -2.1920 \\
    Kurtosis & 107.5232 & 70.1941 & 85.9471 & 45.6609 & 53.0422 & 61.2545 & 27.2200 & 52.9933 \\
    Jarque-Bera & 2775514 & 1003188 & 1292625 & 320676 & 384041 & 460864 & 83211 & 287323 \\
    \bottomrule
    \end{tabular}%
    }
  \label{tab:returnstat}%
\end{table}%

The analysis of the long range dependence is rather similar for the different time scales. The Hurst exponent profiles for the different subsamples are close regarding temporal behavior and range. In all cases, we notice a marked persistent (procyclical) behavior until 2014. After such year, the time series of Hurst exponents seem to stabilize around a value of $0.5\pm0.05$, inducing to think in a more informational efficient market. However we cannot find the reason for such change in the dynamics, giving the unconnectedness of price behavior with market fundamentals.  

\begin{figure}[h!p]
    \centering
        \subfloat[5-hour return]{%
            \includegraphics[scale=0.4]{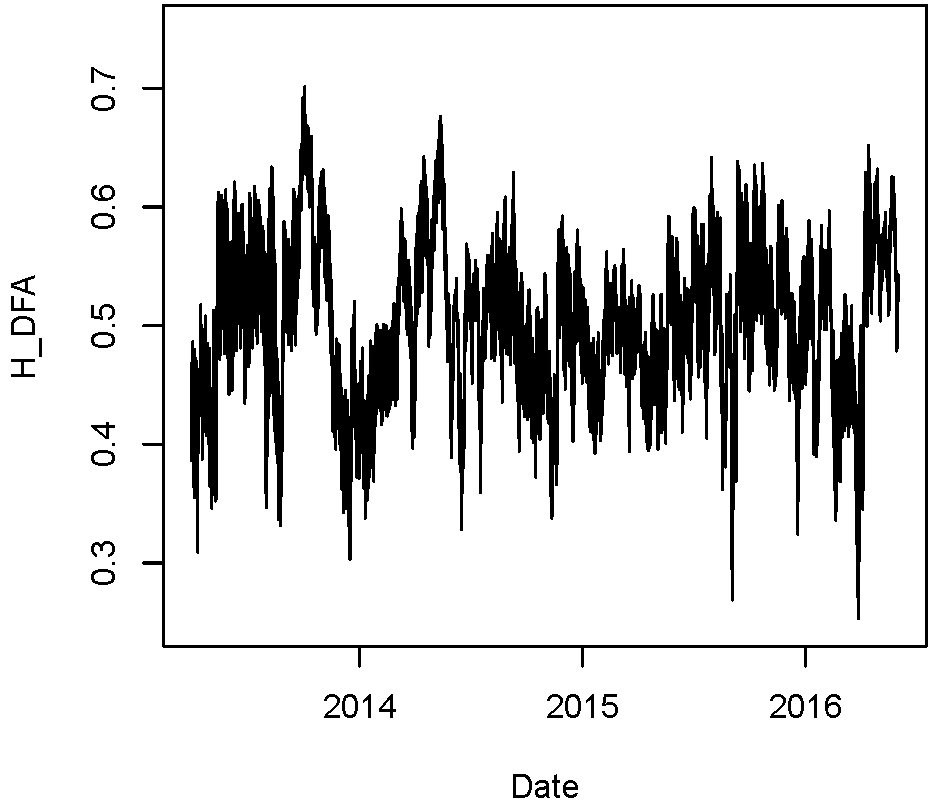}
            }%
        \subfloat[6-hour return]{%
           \includegraphics[scale=0.4]{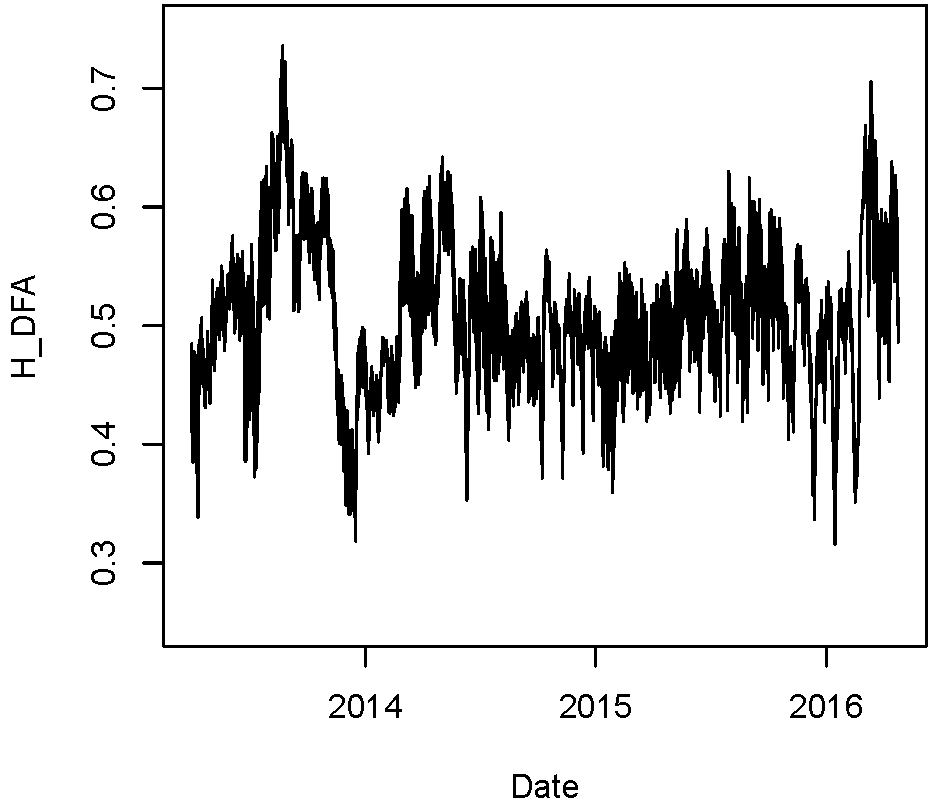}
        } \\
        \subfloat[7-hour return]{%
            \includegraphics[scale=0.4]{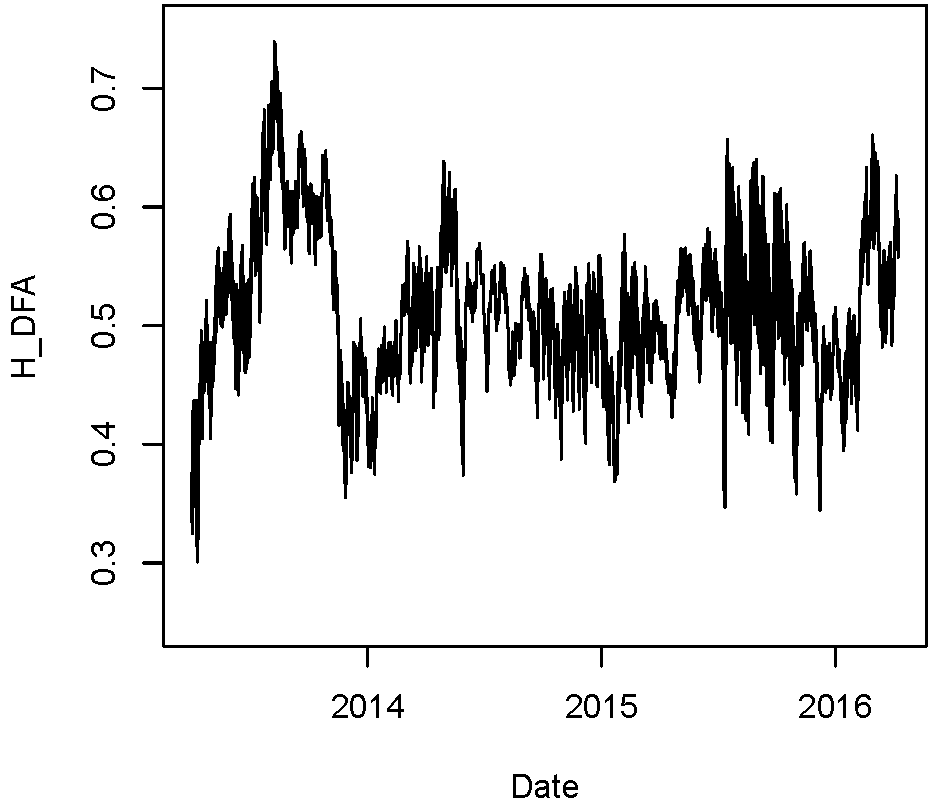}
            }%
        \subfloat[8-hour return]{%
           \includegraphics[scale=0.4]{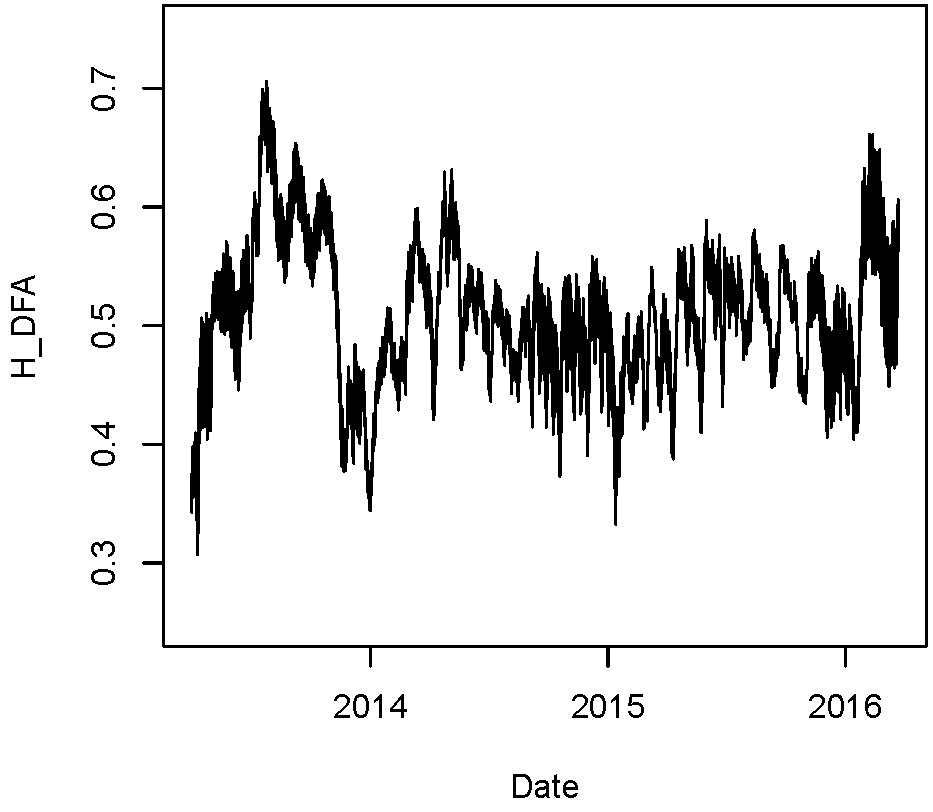}
        } \\
				 \subfloat[9-hour return]{%
           \includegraphics[scale=0.4]{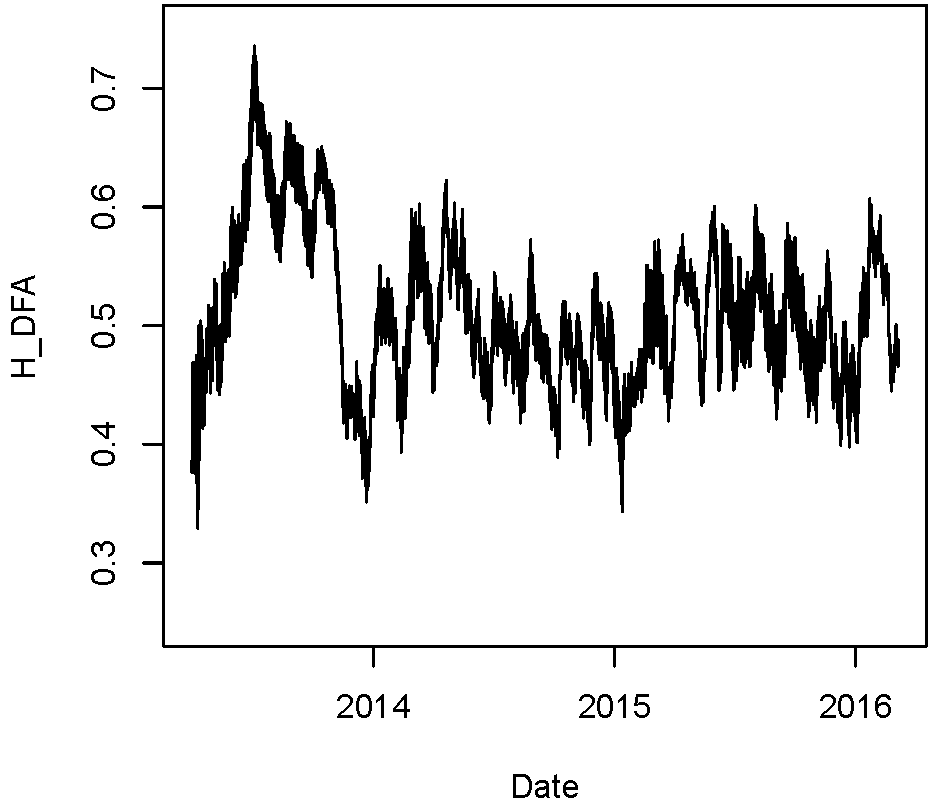}
        }
				 \subfloat[10-hour return]{%
           \includegraphics[scale=0.4]{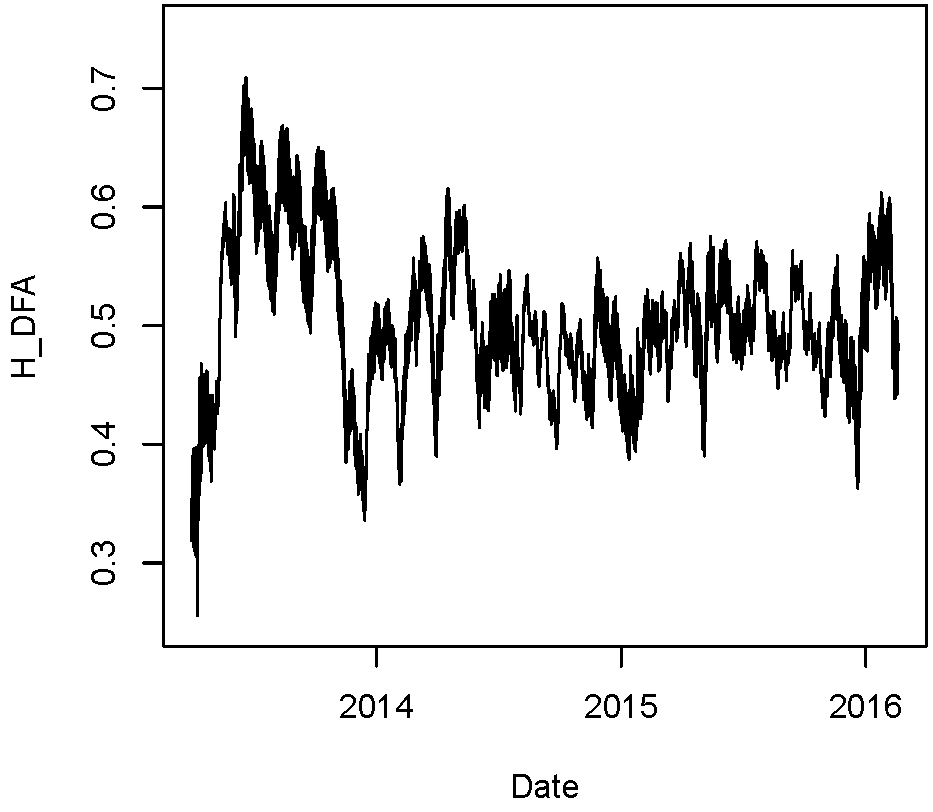}
        } \\
				 \subfloat[11-hour return]{%
           \includegraphics[scale=0.4]{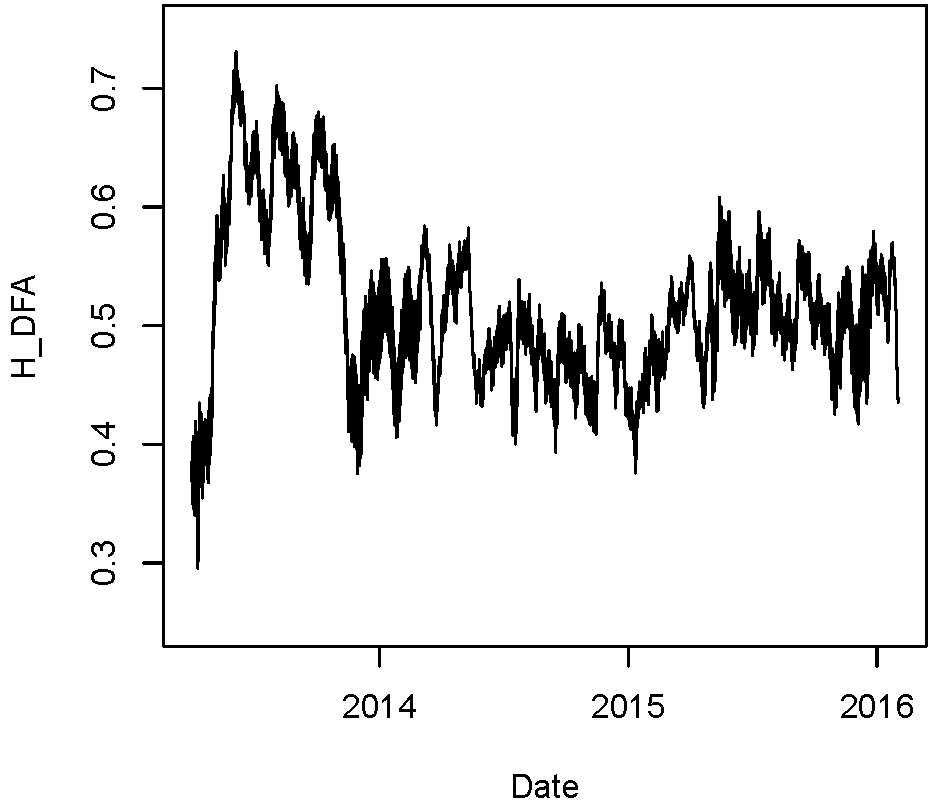}
        }
				 \subfloat[12-hour return]{%
           \includegraphics[scale=0.4]{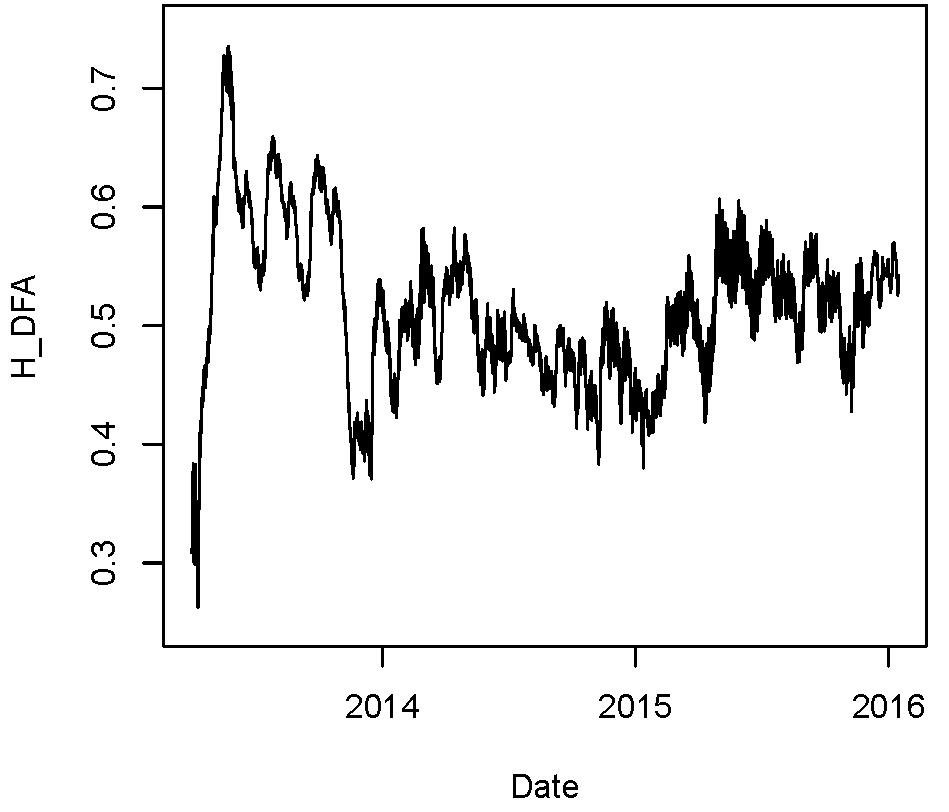}
        }
				\caption{Hurst exponent using DFA method, for 5 to 12 hour BTC returns, using a sliding window of 500 datapoints and one datapoint step forward. Period: 2013-2016}
   \label{fig:dfa-hour-return}
			\end{figure}

\section{Conclusions \label{sec:conclusions}}

In this paper we study the long range memory and other statistical properties of Bitcoin daily and intraday prices. The period under study goes from 2011 until 2017. We compute the Hurst exponent by means of the Detrended Fluctuation Analysis method, using a sliding window in order to assess the time-varying long range dependence. We detect that:
\begin{enumerate}
\item In spite of the fact that Bitcoin presents large volatility, it is reducing over time.
\item We find that the long range memory is not related to market liquidity.
\item The behavior across different time scales (5 to 12 hours) is essentially similar, in terms of long range memory.
\item Until 2014 the time series had a persistent behavior ($H>0.5$), whereas after such date, the Hurst exponent tended to move around 0.5.
\end{enumerate} 
In light of our results, more research should be done in order to uncover the reason for the change in Bitcoin dynamics across time.

\section*{\refname}
\bibliographystyle{apalike}
\bibliography{corporatephysics2}   

\begin{thebibliography}{}

\bibitem[Ali et~al., 2014]{BoEngland}
Ali, R., Barrdear, J., Clews, R., and Southgate, J. (2014).
\newblock The economics of digital currencies.
\newblock {\em Bank of England Quarterly Bulletin}, 54(3):276--286.

\bibitem[Bachelier, 1900]{Bachel}
Bachelier, L. (1900).
\newblock {\em Th\'eorie de la sp\'eculation}.
\newblock Annales scientifiques de l'\'Ecole Normale Sup\'erieure, Paris.

\bibitem[Bariviera et~al., 2015a]{EPJB2015}
Bariviera, A., Guercio, M., Martinez, L., and Rosso, O. (2015a).
\newblock The (in)visible hand in the libor market: an information theory
  approach.
\newblock {\em The European Physical Journal {B}}, 88(8):208.

\bibitem[Bariviera et~al., 2015b]{BarivieraRSTA2015}
Bariviera, A., Guercio, M., Martinez, L., and Rosso, O. (2015b).
\newblock A permutation information theory tour through different interest rate
  maturities: the libor case.
\newblock {\em Philosophical Transactions of the Royal Society of London A:
  Mathematical, Physical and Engineering Sciences}, 373:20150119.

\bibitem[Bariviera, 2011]{Bariviera2011}
Bariviera, A.~F. (2011).
\newblock The influence of liquidity on informational efficiency: The case of
  the thai stock market.
\newblock {\em Physica A: Statistical Mechanics and its Applications},
  390(23-24):4426--4432.

\bibitem[Bariviera et~al., 2012]{BaGuMa12}
Bariviera, A.~F., Guercio, M.~B., and Martinez, L.~B. (2012).
\newblock A comparative analysis of the informational efficiency of the fixed
  income market in seven european countries.
\newblock {\em Economics Letters}, 116(3):426--428.

\bibitem[Bariviera et~al., 2014]{BaGuMa14}
Bariviera, A.~F., Guercio, M.~B., and Martinez, L.~B. (2014).
\newblock Informational efficiency in distressed markets: The case of european
  corporate bonds.
\newblock {\em The Economic and Social Review}, 45(3):349--369.

\bibitem[Barkoulas and Baum, 1996]{BarkoulasBaum96}
Barkoulas, J.~T. and Baum, C.~F. (1996).
\newblock Long-term dependence in stock returns.
\newblock {\em Economics Letters}, 53(3):253--259.

\bibitem[Barkoulas et~al., 2000]{BarkoulasBaumTravlos00}
Barkoulas, J.~T., Baum, C.~F., and Travlos, N. (2000).
\newblock Long memory in the greek stock market.
\newblock {\em Applied Financial Economics}, 10(2):177--184.

\bibitem[BitcoinCharts, 2016]{bitcoincharts}
BitcoinCharts (2016).
\newblock Bitcoin charts.
\newblock \url{http://bitcoincharts.com/}.
\newblock Accessed: 2016-12-27.

\bibitem[Blasco and Santamar\'ia, 1996]{BlascoSantamaria96}
Blasco, N. and Santamar\'ia, R. (1996).
\newblock Testing memory patterns in the spanish stock market.
\newblock {\em Applied Financial Economics}, 6(5):401--411.

\bibitem[Cajueiro and Tabak, 2010]{CajueiroTabak2010}
Cajueiro, D. and Tabak, B. (2010).
\newblock Fluctuation dynamics in us interest rates and the role of monetary
  policy.
\newblock {\em Finance Research Letters}, 7(3):163--169.

\bibitem[Cajueiro and Tabak, 2004a]{CajuTab}
Cajueiro, D.~O. and Tabak, B.~M. (2004a).
\newblock Evidence of long range dependence in asian equity markets: the role
  of liquidity and market restrictions.
\newblock {\em Physica A: Statistical Mechanics and its Applications},
  342(3-4):656--664.

\bibitem[Cajueiro and Tabak, 2004b]{CajuTabEM}
Cajueiro, D.~O. and Tabak, B.~M. (2004b).
\newblock The hurst exponent over time: testing the assertion that emerging
  markets are becoming more efficient.
\newblock {\em Physica A: Statistical and Theoretical Physics},
  336(3-4):521--537.

\bibitem[Cajueiro and Tabak, 2004c]{CajueiroTabak04}
Cajueiro, D.~O. and Tabak, B.~M. (2004c).
\newblock Ranking efficiency for emerging markets.
\newblock {\em Chaos, Solitons and Fractals}, 22(2):349--352.

\bibitem[Cajueiro and Tabak, 2005]{CajueiroTabak05causes}
Cajueiro, D.~O. and Tabak, B.~M. (2005).
\newblock Possible causes of long-range dependence in the brazilian stock
  market.
\newblock {\em Physica A: Statistical Mechanics and its Applications},
  345(3-4):635--645.

\bibitem[Cajueiro and Tabak, 2007]{CajueiroTabakUS}
Cajueiro, D.~O. and Tabak, B.~M. (2007).
\newblock Time-varying long-range dependence in us interest rates.
\newblock {\em Chaos, Solitons \& Fractals}, 34(2):360 -- 367.

\bibitem[Cajueiro and Tabak, 2009]{CajueiroTabakTSIR}
Cajueiro, D.~O. and Tabak, B.~M. (2009).
\newblock Testing for long-range dependence in the brazilian term structure of
  interest rates.
\newblock {\em Chaos, Solitons \& Fractals}, 40(4):1559 -- 1573.

\bibitem[Carbone et~al., 2004]{Carbone04}
Carbone, A., Castelli, G., and Stanley, H.~E. (2004).
\newblock Time-dependent hurst exponent in financial time series.
\newblock {\em Physica A: Statistical Mechanics and its Applications},
  344(1-2):267--271.

\bibitem[Cheong, 2010]{Cheong2010}
Cheong, C. (2010).
\newblock Estimating the hurst parameter in financial time series via heuristic
  approaches.
\newblock {\em Journal of Applied Statistics}, 37(2):201--214.

\bibitem[Cheung et~al., 2015]{CheungRoca2015}
Cheung, A. W.-K., Roca, E., and Su, J.-J. (2015).
\newblock {Crypto-currency bubbles: an application of the Phillips?Shi?Yu
  (2013) methodology on Mt. Gox bitcoin prices}.
\newblock {\em Applied Economics}, 47(23):2348--2358.

\bibitem[Cheung and Lai, 1995]{CheungLai95}
Cheung, Y.~. and Lai, K.~S. (1995).
\newblock A search for long memory in international stock market returns.
\newblock {\em Journal of International Money and Finance}, 14(4):597--615.

\bibitem[Ciaian et~al., 2016]{Ciaian2016}
Ciaian, P., Rajcaniova, M., and d'Artis Kancs (2016).
\newblock {The economics of BitCoin price formation}.
\newblock {\em Applied Economics}, 48(19):1799--1815.

\bibitem[Coinmap, 2016]{coinmap}
Coinmap (2016).
\newblock Coinmap 2.0.
\newblock \url{https://coinmap.org/#/world/41.11246879/1.40625000/2}.
\newblock Accessed: 2016-12-27.

\bibitem[Coinmarket, 2016]{coinmarketcap}
Coinmarket (2016).
\newblock {Crypto-Currency Market Capitalizations}.
\newblock \url{https://coinmarketcap.com/currencies/}.
\newblock Accessed: 2016-12-27.

\bibitem[Fama, 1970]{Fama70}
Fama, E.~F. (1970).
\newblock Efficient capital markets: A review of theory and empirical work.
\newblock {\em The Journal of Finance}, 25(2, Papers and Proceedings of the
  Twenty-Eighth Annual Meeting of the American Finance Association New York,
  N.Y. December, 28-30, 1969):pp. 383--417.

\bibitem[Fama, 1976]{Fama76}
Fama, E.~F. (1976).
\newblock {\em Foundations of finance : portfolio decisions and securities
  prices}.
\newblock Basic Books, New York.

\bibitem[Fama and French, 1988]{FF88}
Fama, E.~F. and French, K.~R. (1988).
\newblock Dividend yields and expected stock returns.
\newblock {\em Journal of Financial Economics}, 22(1):3--25.

\bibitem[Feigenbaum and Freund, 1996]{FeigenbaumFreund1996}
Feigenbaum, J.~A. and Freund, P. G.~O. (1996).
\newblock Discrete scale invariance in stock markets before crashes.
\newblock {\em International Journal of Modern Physics B}, 10(27):3737--3745.

\bibitem[Geweke and Porter-Hudak, 1983]{GPH83}
Geweke, J. and Porter-Hudak, S. (1983).
\newblock The estimation and application of long memory time series models.
\newblock {\em Journal of Time Series Analysis}, 4:221--238.

\bibitem[Grau-Carles, 2000]{GrauCarles}
Grau-Carles, P. (2000).
\newblock Empirical evidence of long-range correlations in stock returns.
\newblock {\em Physica A: Statistical Mechanics and its Applications},
  287(3-4):396--404.

\bibitem[Grau-Carles, 2005]{GrauCarles05}
Grau-Carles, P. (2005).
\newblock Tests of long memory: A bootstrap approach.
\newblock {\em Computational Economics}, 25(1-2):103--113.

\bibitem[Greene and Fielitz, 1977]{GreeneFielitz77}
Greene, M.~T. and Fielitz, B.~D. (1977).
\newblock Long-term dependence in common stock returns.
\newblock {\em Journal of Financial Economics}, 4(3):339--349.

\bibitem[Henry, 2002]{Henry02}
Henry, Ólan, T. (2002).
\newblock Long memory in stock returns: Some international evidence.
\newblock {\em Applied Financial Economics}, 12(10):725--729.

\bibitem[Jarque and Bera, 1987]{JarqueBera1987}
Jarque, C.~M. and Bera, A.~K. (1987).
\newblock {A Test for Normality of Observations and Regression Residuals}.
\newblock {\em International Statistical Review / Revue Internationale de
  Statistique}, 55(2):163--172.

\bibitem[Kasman et~al., 2009]{KasmanTurgutAyhan09}
Kasman, S., Turgutlu, E., and Ayhan, A.~D. (2009).
\newblock Long memory in stock returns: Evidence from the major emerging
  central european stock markets.
\newblock {\em Applied Economics Letters}, 16(17):1763--1768.

\bibitem[Kilic, 2004]{Kilic04}
Kilic, R. (2004).
\newblock On the long memory properties of emerging capital markets: Evidence
  from istanbul stock exchange.
\newblock {\em Applied Financial Economics}, 14(13):915--922.

\bibitem[Kristoufek and Vosvrda, 2016]{Kristoufek201627}
Kristoufek, L. and Vosvrda, M. (2016).
\newblock Gold, currencies and market efficiency.
\newblock {\em Physica A: Statistical Mechanics and its Applications}, 449:27
  -- 34.

\bibitem[La~Spada et~al., 2008]{LaSpada08}
La~Spada, G., Farmer, J., and Lillo, F. (2008).
\newblock The non-random walk of stock prices: the long-term correlation
  between signs and sizes.
\newblock {\em The European Physical Journal B - Condensed Matter and Complex
  Systems}, 64(3):607 -- 614.

\bibitem[Laursen and Kyed, 2014]{VCdenmark}
Laursen, A. and Kyed, J.~H. (2014).
\newblock Virtual currencies.
\newblock {\em Danmarks Nationalbank Monetary Review}, 1st. Quarter:85--90.

\bibitem[Lo, 1991]{Lo91}
Lo, A. (1991).
\newblock Long-term memory in stock market prices.
\newblock {\em Econometrica}, 59:1279--1313.

\bibitem[Mills, 1993]{Mills93}
Mills, T.~C. (1993).
\newblock Is there long-term memory in uk stock returns?
\newblock {\em Applied Financial Economics}, 3:303--306.

\bibitem[Montanari et~al., 1999]{Montanari99}
Montanari, A., Taqqu, M.~S., and Teverovsky, V. (1999).
\newblock Estimating long-range dependence in the presence of periodicity: An
  empirical study.
\newblock {\em Mathematical and Computer Modelling}, 29(10-12):217--228.

\bibitem[Murphy et~al., 2015]{Congressional}
Murphy, E.~V., Murphy, M.~M., and Seitzinger, M.~V. (2015).
\newblock Bitcoin: Questions, answers, and analysis of legal issues.
\newblock {CRS} Report 7-5700, Congressional Research Service.

\bibitem[Nakamoto, 2009]{Nakamoto}
Nakamoto, S. (2009).
\newblock Bitcoin: A peer-to-peer electronic cash system.
\newblock \url{https://bitcoin.org/bitcoin.pdf/}.
\newblock Accessed: 2016-12-27.

\bibitem[Panas, 2001]{Panas01}
Panas, E. (2001).
\newblock Estimating fractal dimension using stable distributions and exploring
  long memory through arfima models in athens stock exchange.
\newblock {\em Applied Financial Economics}, 11(4):395--402.

\bibitem[Peng et~al., 1994]{Mosaic94}
Peng, C.-K., Buldyrev, S.~V., Havlin, S., Simons, M., Stanley, H.~E., and
  Goldberger, A.~L. (1994).
\newblock Mosaic organization of dna nucleotides.
\newblock {\em Physical Review E}, 49(2):1685--1689.

\bibitem[Peng et~al., 1995]{Peng95}
Peng, C.-K., Havlin, S., Stanley, H.~E., and Goldberger, A.~L. (1995).
\newblock Quantification of scaling exponents and crossover phenomena in
  nonstationary heartbeat time series.
\newblock {\em Chaos: An Interdisciplinary Journal of Nonlinear Science},
  5(1):82--87.

\bibitem[Rosso et~al., 2007]{Rosso07}
Rosso, O.~A., Larrondo, H.~A., Martin, M.~T., Plastino, A., and Fuentes, M.~A.
  (2007).
\newblock Distinguishing noise from chaos.
\newblock {\em Phys. Rev. Lett.}, 99(15):154102.

\bibitem[Serinaldi, 2010]{Serinaldi10}
Serinaldi, F. (2010).
\newblock Use and misuse of some hurst parameter estimators applied to
  stationary and non-stationary financial time series.
\newblock {\em Physica A: Statistical Mechanics and its Applications},
  389(14):2770--2781.

\bibitem[Taqqu et~al., 1995]{Taqqu95}
Taqqu, M.~S., Teverovsky, V., and Willinger, W. (1995).
\newblock Estimators for long-range dependence: An empirical study.
\newblock {\em Fractals}, 3:785--798.

\bibitem[Tolvi, 2003]{Tolvi03}
Tolvi, J. (2003).
\newblock Long memory and outliers in stock market returns.
\newblock {\em Applied Financial Economics}, 13(7):495--502.

\bibitem[Ureche-Rangau and de~Rorthays, 2009]{UrecheRangau09}
Ureche-Rangau, L. and de~Rorthays, Q. (2009).
\newblock More on the volatility-trading volume relationship in emerging
  markets: The chinese stock market.
\newblock {\em Journal of Applied Statistics}, 36(7):779--799.

\bibitem[Vodenska-Chitkushev et~al., 2008]{Vodenskaetal08}
Vodenska-Chitkushev, I., Wang, F.~Z., Weber, P., Yamasaki, K., Havlin, S., and
  Stanley, H.~E. (2008).
\newblock Comparison between volatility return intervals of the s \& p 500
  index and two common models.
\newblock {\em The European Physical Journal B - Condensed Matter and Complex
  Systems}, 61(2):217 -- 223.

\bibitem[Wright, 2001]{Wright01}
Wright, J. (2001).
\newblock Long memory in emerging market stock returns.
\newblock {\em Emerging Markets Quarterly}, 5:50--55.

\bibitem[Zunino et~al., 2012]{Zunino2012}
Zunino, L., Bariviera, A.~F., Guercio, M.~B., Martinez, L.~B., and Rosso, O.~A.
  (2012).
\newblock On the efficiency of sovereign bond markets.
\newblock {\em Physica A: Statistical Mechanics and its Applications},
  391(18):4342--4349.

\bibitem[Zunino et~al., 2007]{ZuninoPhyB07}
Zunino, L., Tabak, B.~M., P\'erez, D.~G., Garavaglia, M., and Rosso, O.~A.
  (2007).
\newblock Inefficiency in latin-american market indices.
\newblock {\em The European Physical Journal B - Condensed Matter and Complex
  Systems}, 60:111--121.

\bibitem[Zunino et~al., 2011]{ZuninoPermutation11}
Zunino, L., Tabak, B.~M., Serinaldi, F., Zanin, M., P\'erez, D.~G., and Rosso,
  O.~A. (2011).
\newblock Commodity predictability analysis with a permutation information
  theory approach.
\newblock {\em Physica A: Statistical Mechanics and its Applications},
  390(5):876--890.

\bibitem[Zunino et~al., 2010]{ZuninoCausality10}
Zunino, L., Zanin, M., Tabak, B.~M., P\'erez, D.~G., and Rosso, O.~A. (2010).
\newblock Complexity-entropy causality plane: A useful approach to quantify the
  stock market inefficiency.
\newblock {\em Physica A: Statistical Mechanics and its Applications},
  389(9):1891--1901.

\end{thebibliography}

\end{document}